\documentclass[aps,prd,showpacs,showkeys,amsmath,amssymb]{revtex4}
\usepackage{amsfonts}
\usepackage{graphicx}
\usepackage{amsmath}
\usepackage{mathrsfs}
\usepackage{tabls}
\usepackage{slashed}
\usepackage{bm}

\usepackage{color}

\usepackage{booktabs}
\usepackage{multirow}
\usepackage{array}

\usepackage{subfig}

\allowdisplaybreaks

\begin{document}
\begin{picture}(5,5)
\put(375,20){{\footnotesize\underline{Submitted to `Chinese Physics C'}}}
\end{picture}

\title{$\rho$ Meson Decays of Heavy Hybrid Mesons}
\author{Liang Zhang}
\affiliation{School of Physics, Beihang University, Beijing 100191, China}
\author{Peng-Zhi Huang}
\email{pzhuang@pku.edu.cn}
\affiliation{Department of Physics
and State Key Laboratory of Nuclear Physics and Technology\\
Peking University, Beijing 100871, China}
\affiliation{Theoretical Physics Center for Science Facilities, CAS, Beijing 100049, China}

\begin{abstract}
We calculate the $\rho$ meson couplings between the heavy hybrid doublets $H^h/S^h/M^h/T^h$
and the ordinary $q\bar{Q}$ doublets in the framework of the light-cone QCD sum rule.
The sum rules obtained rely mildly on the Borel parameters
in their working regions.
The resulting coupling constants are rather small in most cases.
\end{abstract}
\keywords{Heavy hybrid meson, QCD sum rule, Heavy quark effective theory}
\pacs{12.39.Mk, 12.38.Lg, 12.39.Hg}
\maketitle
\pagenumbering{arabic}

\section{Introduction}\label{introduction}

Hadron states that do not fit into the constituent quark model have been studied widely in the past several
decades.
In recent years, the discovery of a number of unexpected exotic resonances such as the so called \textit{XYZ} mesons
has revitalized the research of the existence of unconventional hadron states and their nature.

Theoretically, Quantum Chromodynamics (QCD), the fundamental theory of the strong interaction, may allow
a far richer spectrum than the conventional quark model.
Fox example, hybrids ($q\bar{q}g,\cdots$), glueballs ($gg$,
$ggg$, $\cdots$), and multi-quark states
($qq\bar{q}\bar{q}$, $qqqq\bar{q}$, $\cdots$) may not be prohibited by QCD.
Those with $J^{PC}=0^{--}, 0^{+-}, 1^{-+}, 2^{+-},\cdots$ are called ``exotic'' states.
They have attracted much interest
because they are not allowed by the constituent quark model
and do not mix with the ordinary mesons.

Evidence of exotic mesons with $J^{PC}=1^{-+}$,
e.g. $\pi_1(1400)$ \cite{1400}, $\pi_1(1600)$ \cite{1600},
have emerged in the last few years.
They are usually considered as candidates of hybrid mesons
and have been studied extensively in various frameworks
such as QCD sum rules, lattice QCD, AdS/QCD, the flux tube model, etc.
The masses and decay properties of the $1^{-+}$ states have been studied in the framework of QCD sum rules \cite{QSRmass,QSRdecay}.

Based on the accumulated evidence of these light hybrid mesons,
it is plausible to assume the existence of
heavy quarkonium hybrids ($Q\bar{Q}g$) and
heavy hybrid mesons containing one heavy quark ($q\bar{Q}g$)
which may be not exotic.
Govaerts et al. have studied these states in several works \cite{Govaerts}.
In \cite{zhu1}, the masses of $Q\bar{Q}g$ were calculated
at the leading order of heavy quark effective theory (HQET) \cite{hqet}.
In \cite{zhu2}, the masses of $q\bar{Q}g$ and their pionic couplings to ordinary heavy mesons
were calculated.

In the heavy quark limit,
the binding energy and the pionic couplings of $q\bar{Q}g$ to $q\bar{Q}$
were worked out in \cite{HeavyHybridparameter}
by Shifman-Vainshtein-Zakharov (SVZ) sum rules \cite{svz}.
HQET describes the large mass ($m_Q$) asymptotics.
At the leading order of this theory, the Lagrangian
is endowed with the heavy quark flavor-spin symmetry,
and the spectrum of $q\bar{Q}$ consists of degenerate doublets.
The components of a doublet share the same $j_l$, the
angular momentum of the light degrees of freedom.
For example,
we denote the doublet $(0^-,1^-)$ as $H$, which consists of two $j_l={1 \over 2}$ $S$-wave $q\bar{Q}$.
Similarly, the $P$-wave doublets $(0^+,1^+)/(1^+,2^+)$ are denoted as $S/T$ and
the $D$-wave doublets $(1^-,2^-)/(2^-,3^-)$ as $M/N$.
We denote the two $j_l={1\over 2}$ $q\bar{Q}g$ doublets with parity $P=+$ and $P=-$ as $S^h$ and $H^h$, respectively.
Similarly, we use $T^h$ and $M^h$ to denote the two $j_l={3\over 2}$ doublets
with positive parity and negative parity, respectively.

In this work,
we adopt the light-cone QCD sum rules (LCQSR) approach \cite{light-cone} to
investigate the $\rho$ meson couplings between $q\bar{Q}g$ and $q\bar{Q}$.
We derive the sum rules for the $\rho$ meson couplings between doublets $D^h$ and $D$ ($D=H/S/T/M$)
in Sec.~2.
The numerical analysis is given in Sec.~3,
followed by a brief conclusion in Sec.~4.
The details of the partial amplitudes of these $\rho$ decay channels are presented in Appendix A.
The light-cone wave functions of the $\rho$ meson
involved in our calculation are listed in Appendix B.

\section{$\rho$ meson couplings}\label{rhocouplings}

The interpolating currents for $H^h$ and $M^h$ adopted in our calculation can be written as
\begin{eqnarray}\label{currentsHM}
&&J^{\dag}_{H^h_0}
=\sqrt{\frac{1}{2}}\bar{h}_vig_s\gamma_5\sigma_t\cdot G q\,,\nonumber\\
&&J^{\dag\alpha}_{H^h_1}
=\sqrt{\frac{1}{2}}\bar{h}_vig_s\gamma^\alpha_t\sigma_t\cdot G q\,,\nonumber\\
&&J^{\dag\alpha}_{M^h_1}
=\bar{h}_vg_s\biggl[3G_t^{\alpha\beta}\gamma_\beta+i\gamma^\alpha_t\sigma_t\cdot G\biggl] q\,,\nonumber\\
&&J^{\dag\alpha_1\alpha_2}_{M^h_2}
=\sqrt{\frac{3}{2}}\bar{h}_vg_s\gamma_5\biggl[G_t^{\alpha_1\beta}\gamma_\beta\gamma_t^{\alpha_2}
+G_t^{\alpha_2\beta}\gamma_\beta\gamma_t^{\alpha_1}
-\frac{2}{3}ig^{\alpha_1\alpha_2}_t\sigma_t\cdot G\biggl] q\,,
\end{eqnarray}
where $G_{\alpha\beta}=G_{\alpha\beta}^n\lambda^n/2$ and $h_v(x)=e^{im_Qv\cdot x}\frac{1+\slashed{v}}{2}Q(x)$.
The subscript $t$ means that the corresponding Lorentz tensor is
perpendicular to $v$,
the 4-velocity of the heavy quark.
$g_t^{\alpha\beta}=g^{\alpha\beta}-v^\alpha v^\beta$.
For any asymmetric tensor $\mathcal{A}^{\alpha_1\alpha_2\cdots\alpha_n}$, we may define
\begin{equation}
\mathcal{A}_t^{\alpha_1\alpha_2\cdots\alpha_n}
  =\mathcal{A}^{\alpha_1\alpha_2\cdots\alpha_n}
  -\sum_{i=1}^n(\mathcal{A}^{\alpha_1\cdots\alpha_{i-1}\alpha\alpha_{i+1}\cdots\alpha_n}v_\alpha)v^{\alpha_i}\,.
\end{equation}
We define the overlapping amplitudes between the these interpolating currents and the corresponding hybrids as
\begin{eqnarray}\label{oaHhMh}
&&\langle 0|J_{H^h_0}(0)|H^h_0(v)\rangle=f_{H^h_0}\,,\nonumber\\
&&\langle 0|J_{H^h_1}^\alpha(0)|H^h_1(v, \lambda)\rangle=f_{H^h_1}\eta_{H^h_1}^\alpha(v, \lambda)\,,\nonumber\\
&&\langle 0|J_{M^h_1}^\alpha(0)|M^h_1(v, \lambda)\rangle=f_{M^h_1}\eta_{M^h_1}^\alpha(v, \lambda)\,,\nonumber\\
&&\langle 0|J_{M^h_2}^{\alpha_1\alpha_2}(0)|M^h_2(v, \lambda)\rangle=f_{M^h_2}\eta_{M^h_2}^{\alpha_1\alpha_2}(v, \lambda)\,,
\end{eqnarray}
where $\eta(v, \lambda)$ denotes the polarization of the heavy hybrid.
These symmetric traceless tensors are perpendicular to $v$, namely $\eta_{\alpha\alpha_2\cdots\alpha_n}v^\alpha=0$.

We obtain the interpolating currents for the doublets $S^h$ and $T^h$
by simply inserting $\gamma_5$ into the currents in Eq.~(\ref{currentsHM}):
\begin{eqnarray}\label{currentsST}
&&J^{\dag}_{S^h_0}
=\sqrt{\frac{1}{2}}\bar{h}_vig_s\sigma_t\cdot G q\,,\nonumber\\
&&J^{\dag\alpha}_{S^h_1}
=\sqrt{\frac{1}{2}}\bar{h}_vig_s\gamma_5\gamma^\alpha_t\sigma_t\cdot G q\,,\nonumber\\
&&J^{\dag\alpha}_{T^h_1}
=\bar{h}_vg_s\gamma_5\biggl[3G_t^{\alpha\beta}\gamma_\beta+i\gamma^\alpha_t\sigma_t\cdot G\biggl] q\,,\nonumber\\
&&J^{\dag\alpha_1\alpha_2}_{T^h_2}
=\sqrt{\frac{3}{2}}\bar{h}_vg_s\biggl[G_t^{\alpha_1\beta}\gamma_\beta\gamma_t^{\alpha_2}
+G_t^{\alpha_2\beta}\gamma_\beta\gamma_t^{\alpha_1}
-\frac{2}{3}ig^{\alpha_1\alpha_2}_t\sigma_t\cdot G\biggl] q\,.
\end{eqnarray}
The corresponding overlapping amplitudes and projection operators
can be defined similarly to Eq.~(\ref{oaHhMh}).

The interpolating currents
for $q\bar{Q}$ doublets $H$ and $S$ read:
\begin{eqnarray}
J^\dag_{H_0}&=&\sqrt{\frac{1}{2}}\bar{h}_v\gamma_5 q\,,\nonumber\\
J^{\dag\alpha}_{H_1}&=&\sqrt{\frac{1}{2}}\bar{h}_v\gamma_t^\alpha q\,,\nonumber\\
J^\dag_{S_0}&=&\sqrt{\frac{1}{2}}\bar{h}_v q\,,\nonumber\\
J^{\dag\alpha}_{S_1}&=&\sqrt{\frac{1}{2}}\bar{h}_v\gamma_t^\alpha\gamma_5 q\,.
\end{eqnarray}
The amplitudes between the ordinary heavy mesons and the states created by these currents acting on the vacuum state
are
\begin{eqnarray}\label{oaHM}
&&\langle 0|J_{H_0}(0)|H_0(v)\rangle=f_{H_0}\,,\nonumber\\
&&\langle 0|J_{H_1}^\alpha(0)|H_1(v, \lambda)\rangle=f_{H_1}\epsilon_{H_1}^\alpha(v, \lambda)\,,\nonumber\\
&&\langle 0|J_{S_0}(0)|S_0(v)\rangle=f_{S_0}\,,\nonumber\\
&&\langle 0|J_{S_1}^\alpha(0)|S_1(v, \lambda)\rangle=f_{S_1}\epsilon_{S_1}^\alpha(v, \lambda)\,.
\end{eqnarray}

Here we outline the deduction of the sum rules for
$g^{p0}_{H^h_1H_1\rho}$ and $g^{p1}_{H^h_1H_1\rho}$,
where $p$ is the orbital angular momentum of the $\rho$ meson, the superscript `0' and `1' are the total angular momentum of the $\rho$ meson.
We define $g^{p0}_{H^h_1H_1\rho}$ and $g^{p1}_{H^h_1H_1\rho}$
in term of the decay amplitude of the process $H^h_1\rightarrow H_1+\rho$:
\begin{eqnarray}
\mathcal {M}(H^h_1\rightarrow H_1+\rho)
&=&I\Big[(e^*\cdot \eta_t)(\epsilon^*\cdot q_t)-(e^*\cdot \epsilon^*_t)(\eta\cdot q_t)\Big]g^{p1}_{H^h_1H_1\rho}
+I(e^*\cdot q_t)(\epsilon^*\cdot \eta_t)g^{p0}_{H^h_1H_1\rho}\,,
\end{eqnarray}
where $\eta$, $\epsilon^*$ and $e^*$ are the polarization of $H^h_1$, $H_1$ and $\rho$, respectively,
and $q$ denotes the momentum of the $\rho$.
For the charged $\rho$ meson, $I=1$,
and $I=1/\sqrt{2}$ if the final $\rho$ meson is neutral.

We consider the following correlation function:
\begin{eqnarray}
i\int dx\ e^{-ik\cdot x}\langle\rho(q)|J_{H_1}^\beta(0)J^{\dag\alpha}_{H^h_1}(x)|0\rangle
=I \left[e_t^\alpha q_t^\beta-q_t^\alpha e_t^\beta\right] G^{p1}_{H^h_1H_1\rho}(\omega,\omega')
+I g_t^{\alpha\beta}(e\cdot q_t) G^{p0}_{H^h_1H_1\rho}(\omega,\omega')\,,
\end{eqnarray}
where $\omega=2k\cdot v$ and $\omega'=2(k-q)\cdot v$,
and we have the following dispersion relation
\begin{eqnarray}\label{dispersionrelation}
G^{p1}_{H^h_1H_1\rho}(\omega,\omega')
=\int_0^\infty ds_1\int_0^\infty ds_2 \frac{\rho^{p1}_{H^h_1H_1\rho}(s_1,s_2)}{(s_1-\omega-i\epsilon)(s_2-\omega'-i\epsilon)}
+\int_0^\infty ds_1\frac{\rho^{p1}_1(s_1)}{s_1-\omega-i\epsilon}
+\int_0^\infty ds_2\frac{\rho^{p1}_2(s_2)}{s_2-\omega'-i\epsilon}
+\cdots\,,~
\end{eqnarray}
with
\begin{eqnarray}
\rho^{p1}_{H^h_1H_1\rho}(s_1,s_2)
=f_{H^h} f_H g^{p1}_{H^h_1H_1\rho}\delta(s_1-2\Lambda_{H^h})\delta(s_2-2\Lambda_H)+\cdots\,.
\end{eqnarray}
The case of $G^{p0}_{H^h_1H_1\rho}(\omega,\omega')$ is similar.
$G^{p1}_{H^h_1H_1\rho}(\omega,\omega')$ can be worked out by OPE near the light-cone when $\omega, \omega' \ll0$,
and be formulated with the $\rho$ meson light-cone wave functions
\begin{eqnarray}
G^{p0}_{H^h_1H_1\rho}(\omega,\omega')&=&-\frac{1}{4}\int_0^\infty dt\int \mathcal {D}\underline{\alpha}\ e^{it(\frac{\bar{u}}{2}\omega+\frac{u}{2}\omega')}
\frac{m^2}{(q\cdot v)^3}\nonumber\\
&&\biggl[-2f_\rho m^3\widetilde\Psi(\underline{\alpha})-2f_\rho^T m^2{\cal T}(\underline{\alpha})(q\cdot v)
+f_\rho m[6\widetilde\Phi(\underline{\alpha})+2\widetilde\Psi(\underline{\alpha})+{\cal A}(\underline{\alpha})](q\cdot v)^2\nonumber\\
&&+2f_\rho^T[{\cal T}(\underline{\alpha})+2{\cal T}_1(\underline{\alpha})+2{\cal T}_2(\underline{\alpha})](q\cdot v)^3\biggr]\,,\nonumber\\
G^{p1}_{H^h_1H_1\rho}(\omega,\omega')&=&\frac{1}{4}\int_0^\infty dt\int \mathcal {D}\underline{\alpha}\ e^{it(\frac{\bar{u}}{2}\omega+\frac{u}{2}\omega')}
\frac{m}{q\cdot v}\nonumber\\
&&\biggl[f_\rho m^2[{\cal V}(\underline{\alpha})+{\cal A}(\underline{\alpha})]
-2f_\rho^T m[{\cal T}_1(\underline{\alpha})-{\cal T}_2(\underline{\alpha})+S(\underline{\alpha})](q\cdot v)
-2f_\rho[{\cal V}(\underline{\alpha})+{\cal A}(\underline{\alpha})](q\cdot v)^2\biggr]\,.\quad
\end{eqnarray}
in which $u\equiv\alpha_2+\alpha_3$ and $\bar{u}\equiv 1-u$.

The double Borel transformation $\mathcal{B}_{\omega}^{T_1}\mathcal{B}_{\omega'}^{T_2}$ eliminates
the terms on the right side of Eq.~(\ref{dispersionrelation}),
except the first one which is a double dispersion relation.
Now we arrive at
\begin{eqnarray}
&&f_{H^h} f_H g^{p0}_{H^h_1H_1\rho} e^{-2\bar{u}_0\Lambda_{H^h}/T-2u_0\Lambda_H/T}\nonumber\\
&&=m_\rho^2
 \biggl\{-{1\over 2}f_\rho m_\rho^3\widetilde\Psi^{\{-3\}}-{1\over 2}f_\rho^T m_\rho^2{\cal T}^{\{-2\}}
-f_\rho m_\rho[6\widetilde\Phi^{[-1]}(u_0)+2\widetilde\Psi^{[-1]}(u_0)+{\cal A}^{[-1]}(u_0)]\nonumber\\
&&\phantom{=}
+f_\rho^T[{\cal T}^{[0]}(u_0)+2{\cal T}_1^{[0]}(u_0)+2{\cal T}_2^{[0]}(u_0)]Tf_0(\frac{\omega'_c}{T})\biggr\}\,,\nonumber
\\
&&f_{H^h} f_H g^{p1}_{H^h_1H_1\rho} e^{-2\bar{u}_0\Lambda_{H^h}/T-2u_0\Lambda_H/T}\nonumber\\
&&=m_\rho
 \biggl\{f_\rho m_\rho^2[{\cal V}^{[-1]}(u_0)+{\cal A}^{[-1]}(u_0)]
+f_\rho^T m_\rho[{\cal T}_1^{[0]}(u_0)-{\cal T}_2^{[0]}(u_0)+S^{[0]}(u_0)]Tf_0(\frac{\omega'_c}{T})\nonumber\\
&&\phantom{=}
-{1\over 2}f_\rho[{\cal V}^{[1]}(u_0)+{\cal A}^{[1]}(u_0)]T^2f_1(\frac{\omega'_c}{T})\biggr\}\,,
\end{eqnarray}
with
\begin{eqnarray}
T=\frac{T_1T_2}{T_1+T_2},~~~~u_0=\frac{T_1}{T_1+T_2},~~~~f_n(x)=1-e^{-x}\sum^n_{i=0}\frac{x^i}{i!}\,.
\end{eqnarray}
Here we employ functions $f_n(x)$ to subtract the contribution of the continuum.

$\mathcal{F}^{[\alpha_i]}$s are defined as
\begin{eqnarray}
\mathcal{F}^{[0]}(u_0)&\equiv&\int_0^{u_0}\mathcal{F}(\bar{u}_0,\alpha_2,u_0-\alpha_2)\,d\alpha_2\,,\nonumber
\\
\mathcal{F}^{[1]}(u_0)&\equiv&\mathcal{F}(\bar{u}_0,u_0,0)
-\int_0^{u_0}d\alpha_2\,\frac{\partial\mathcal{F}(1-\alpha_2-\alpha_3,\alpha_2,\alpha_3)}{\partial\alpha_3}
\bigg|_{\alpha_3=u_0-\alpha_2}\,,\nonumber
\\
\mathcal{F}^{[2]}(u_0)&\equiv&
\frac{\partial\mathcal{F}(1-\alpha_2,\alpha_2,0)}{\partial\alpha_2}\bigg|_{\alpha_2=u_0}
+\frac{\partial\mathcal{F}(\bar{u}_0-\alpha_3,u_0,\alpha_3)}{\partial\alpha_3}\bigg|_{\alpha_3=0}
-\int_0^{u_0}d\alpha_2\,
\frac{\partial^2\mathcal{F}(1-\alpha_2-\alpha_3,\alpha_2,\alpha_3)}{\partial\alpha_3^2}\bigg|_{\alpha_3=u_0-\alpha_2}\,,\nonumber
\\
\mathcal {F}^{[-1]}(u_0)&\equiv&
\int_0^1\int_0^{1-\alpha_2}\mathcal {F}(1-\alpha_2-\alpha_3,\alpha_2,\alpha_3)d\alpha_3d\alpha_2
-\int_0^{u_0}\int_0^{u_0-\alpha_2}\mathcal {F}(1-\alpha_2-\alpha_3,\alpha_2,\alpha_3)d\alpha_3d\alpha_2\,,\nonumber
\\
\mathcal {F}^{[-2]}(u_0)&\equiv&
\int_0^{1}\int_0^{1-\alpha_2}\int_0^{\alpha_3}\mathcal {F}(1-\alpha_2-x,\alpha_2,x)dxd\alpha_3d\alpha_2
-\int_0^{u_0}\int_0^{u_0-\alpha_2}\int_0^{\alpha_3}\mathcal {F}(1-\alpha_2-x,\alpha_2,x)dxd\alpha_3d\alpha_2\nonumber\\
&&-\bar{u}_0\int_0^1\int_0^{1-\alpha_2}\mathcal {F}(1-\alpha_2-\alpha_3,\alpha_2,\alpha_3)d\alpha_3d\alpha_2\,.
\end{eqnarray}

Using the above mentioned method, we obtain the sum rules of other $\rho$ meson coupling constants as follows. Their definitions are presented in Appendix A.
\begin{eqnarray}
&&f_{H^h} f_S g^{s1}_{H^h_1S_1\rho} e^{-2\bar{u}_0\Lambda_{H^h}/T-2u_0\Lambda_S/T}\nonumber\\
&&={1\over 6}m_\rho
 \biggl\{{1\over 2}f_\rho m_\rho^4[4\Phi-2\widetilde\Phi-2\widetilde\Psi-{\cal A}]^{\{-2\}}\nonumber\\
&&\phantom{=}+4f_\rho^T m_\rho^3[{\cal T}^{[-1]}(u_0)-2{\cal T}_1^{[-1]}(u_0)+2{\cal T}_2^{[-1]}(u_0)
-2{\cal T}_4^{[-1]}(u_0)+\widetilde S^{[-1]}(u_0)]\nonumber\\
&&\phantom{=}+f_\rho m_\rho^2[-4{\cal V}^{[0]}(u_0)-4\Phi^{[0]}(u_0)+2\widetilde\Phi^{[0]}(u_0)
+2\widetilde\Psi^{[0]}(u_0)-3{\cal A}^{[0]}(u_0)]Tf_0(\frac{\omega'_c}{T})\nonumber\\
&&\phantom{=}-f_\rho^T m_\rho[{\cal T}^{[1]}(u_0)+4{\cal T}_1^{[1]}(u_0)+2{\cal T}_2^{[1]}(u_0)
-2{\cal T}_4^{[1]}(u_0)-2\widetilde S^{[1]}(u_0)]T^2f_1(\frac{\omega'_c}{T})
-f_\rho[{\cal V}^{[2]}(u_0)+{\cal A}^{[2]}(u_0)]T^3f_2(\frac{\omega'_c}{T})\biggr\}\,,\nonumber
\\
&&f_{H^h} f_S g^{d1}_{H^h_1S_1\rho} e^{-2\bar{u}_0\Lambda_{H^h}/T-2u_0\Lambda_S/T}\nonumber\\
&&={1\over 4}m_\rho
 \biggl\{f_\rho m_\rho^2[4\Phi-2\widetilde\Phi-2\widetilde\Psi-{\cal A}]^{\{-2\}}\nonumber\\
&&\phantom{=}+8f_\rho^T m_\rho[{\cal T}^{[-1]}(u_0)+{\cal T}_1^{[-1]}(u_0)+2{\cal T}_2^{[-1]}(u_0)
+{\cal T}_4^{[-1]}(u_0)+\widetilde S^{[-1]}(u_0)]
+4f_\rho[{\cal V}^{[0]}(u_0)+{\cal A}^{[0]}(u_0)]Tf_0(\frac{\omega'_c}{T})\biggr\}\,,\nonumber
\\
&&f_{M^h} f_H g^{p1}_{M^h_1H_1\rho} e^{-2\bar{u}_0\Lambda_{M^h}/T-2u_0\Lambda_H/T}\nonumber\\
&&=\frac{1}{4\sqrt{2}} m_\rho\biggl\{2f_\rho m_\rho^2[{\cal A}^{[-1]}(u_0)-2{\cal V}^{[-1]}(u_0)]\nonumber\\
&&\phantom{=}-2f_\rho^T m_\rho[{\cal T}_2^{[0]}(u_0)-{\cal T}_1^{[0]}(u_0)+2S^{[0]}(u_0)]Tf_0(\frac{\omega'_c}{T})
-f_\rho[{\cal A}_2^{[1]}(u_0)-2{\cal V}^{[1]}(u_0)]T^2f_1(\frac{\omega'_c}{T})\biggl\}\,,\nonumber
\\
&&f_{M^h} f_H g^{p2}_{M^h_1H_1\rho} e^{-2\bar{u}_0\Lambda_{M^h}/T-2u_0\Lambda_H/T}\nonumber\\
&&=-\frac{3}{40\sqrt{2}} m_\rho\biggl\{4f_\rho m_\rho^4\widetilde\Psi^{\{-3\}}
-2f_\rho^T m_\rho^3{\cal T}^{\{-2\}}-4f_\rho m_\rho^2[{\cal A}^{[-1]}(u_0)-4\widetilde\Psi^{[-1]}(u_0)]\nonumber\\
&&\phantom{=}+4f_\rho^Tm_\rho[{\cal T}^{[0]}(u_0)+5{\cal T}_1^{[0]}(u_0)+5{\cal T}_2^{[0]}(u_0)]Tf_0(\frac{\omega'_c}{T})
+6f_\rho{\cal A}^{[1]}(u_0)T^2f_1(\frac{\omega'_c}{T})\biggl\}\,,\nonumber
\\
&&f_{M^h} f_H g^{f2}_{M^h_1H_1\rho} e^{-2\bar{u}_0\Lambda_{M^h}/T-2u_0\Lambda_H/T}\nonumber\\
&&=-\frac{3}{4\sqrt{2}} m_\rho
\biggl\{2f_\rho m_\rho^2\widetilde\Psi^{\{-3\}}-f_\rho^T m_\rho{\cal T}^{\{-2\}}
+8f_\rho {\cal A}^{[-1]}(u_0)\biggl\}\,,\nonumber
\\
&&f_{M^h} f_S g^{s1}_{M^h_1S_1\rho} e^{-2\bar{u}_0\Lambda_{M^h}/T-2u_0\Lambda_S/T}\nonumber\\
&&=\frac{1}{12\sqrt{2}} m_\rho
\biggl\{f_\rho m_\rho^4[2\Phi+2\widetilde\Phi+2\widetilde\Psi+{\cal A}]^{\{-2\}}\nonumber\\
&&\phantom{=}+4f_\rho^T m_\rho^3[{\cal T}^{[-1]}(u_0)-2{\cal T}_1^{[-1]}(u_0)+2{\cal T}_2^{[-1]}(u_0)-2{\cal T}_4^{[-1]}(u_0)-2\widetilde S^{[-1]}(u_0)]\nonumber\\
&&\phantom{=}-2f_\rho m_\rho^2[2{\cal V}^{[0]}(u_0)+2\Phi^{[0]}(u_0)+2\widetilde\Phi^{[0]}(u_0)+2\widetilde\Psi^{[0]}(u_0)-3{\cal A}^{[0]}(u_0)]Tf_0(\frac{\omega'_c}{T})\nonumber\\
&&\phantom{=}-f_\rho^T m_\rho[{\cal T}^{[1]}(u_0)+4{\cal T}_1^{[1]}(u_0)+2{\cal T}_2^{[1]}(u_0)-2{\cal T}_4^{[1]}(u_0)+4\widetilde S^{[1]}(u_0)]T^2f_1(\frac{\omega'_c}{T})
-f_\rho[{\cal V}^{[2]}(u_0)-2{\cal A}^{[2]}(u_0)]T^3f_2(\frac{\omega'_c}{T})\biggl\}\,,\nonumber
\\
&&f_{M^h} f_S g^{d1}_{M^h_1S_1\rho} e^{-2\bar{u}_0\Lambda_{M^h}/T-2u_0\Lambda_S/T}\nonumber\\
&&=-\frac{1}{4\sqrt{2}} m_\rho
\biggl\{f_\rho m_\rho^2[2\Phi+2\widetilde\Phi+2\widetilde\Psi+{\cal A}]^{\{-2\}}\nonumber\\
&&\phantom{=}+4f_\rho^T m_\rho[{\cal T}^{[-1]}(u_0)+2{\cal T}_1^{[-1]}(u_0)+2{\cal T}_2^{[-1]}(u_0)
+{\cal T}_4^{[-1]}(u_0)-2\widetilde S^{[-1]}(u_0)]
+2f_\rho [{\cal V}^{[0]}(u_0)-2{\cal A}^{[0]}(u_0)]Tf_0(\frac{\omega'_c}{T})\biggl\}\,,\nonumber
\\
&&f_{M^h} f_S g^{d2}_{M^h_1S_1\rho} e^{-2\bar{u}_0\Lambda_{M^h}/T-2u_0\Lambda_S/T}\nonumber\\
&&=\frac{3}{2\sqrt{2}} m_\rho
\biggl\{2f_\rho^Tm_\rho[{\cal T}_1^{[-1]}(u_0)+{\cal T}_4^{[-1]}(u_0)]+f_\rho {\cal V}^{[0]}(u_0)Tf_0(\frac{\omega'_c}{T})\biggl\}\,,\nonumber
\\
&&f_{S^h} f_H g^{s1}_{S^h_1H_1\rho} e^{-2\bar{u}_0\Lambda_{S^h}/T-2u_0\Lambda_H/T}\nonumber\\
&&=-{1\over 6}m_\rho
 \biggl\{{1\over 2}f_\rho m_\rho^4[4\Phi-2\widetilde\Phi-2\widetilde\Psi-{\cal A}]^{\{-2\}}\nonumber\\
&&\phantom{=}-4f_\rho^T m_\rho^3[{\cal T}^{[-1]}(u_0)-2{\cal T}_1^{[-1]}(u_0)+2{\cal T}_2^{[-1]}(u_0)
-2{\cal T}_4^{[-1]}(u_0)+\widetilde S^{[-1]}(u_0)]\nonumber\\
&&\phantom{=}+f_\rho m_\rho^2[-4{\cal V}^{[0]}(u_0)-4\Phi^{[0]}(u_0)+2\widetilde\Phi^{[0]}(u_0)
+2\widetilde\Psi^{[0]}(u_0)-3{\cal A}^{[0]}(u_0)]Tf_0(\frac{\omega'_c}{T})\nonumber\\
&&\phantom{=}+f_\rho^T m_\rho[{\cal T}^{[1]}(u_0)+4{\cal T}_1^{[1]}(u_0)+2{\cal T}_2^{[1]}(u_0)
-2{\cal T}_4^{[1]}(u_0)-2\widetilde S^{[1]}(u_0)]T^2f_1(\frac{\omega'_c}{T})
-f_\rho[{\cal V}^{[2]}(u_0)+{\cal A}^{[2]}(u_0)]T^3f_2(\frac{\omega'_c}{T})\biggr\}\,,\nonumber
\\
&&f_{S^h} f_H g^{d1}_{S^h_1H_1\rho} e^{-2\bar{u}_0\Lambda_{S^h}/T-2u_0\Lambda_H/T}\nonumber\\
&&=-{1\over 4}m_\rho
 \biggl\{f_\rho m_\rho^2[4\Phi-2\widetilde\Phi-2\widetilde\Psi-{\cal A}]^{\{-2\}}\nonumber\\
&&\phantom{=}-8f_\rho^T m_\rho[{\cal T}^{[-1]}(u_0)+{\cal T}_1^{[-1]}(u_0)+2{\cal T}_2^{[-1]}(u_0)
+{\cal T}_4^{[-1]}(u_0)+\widetilde S^{[-1]}(u_0)]
+4f_\rho[{\cal V}^{[0]}(u_0)+{\cal A}^{[0]}(u_0)]Tf_0(\frac{\omega'_c}{T})\biggr\}\,,\nonumber
\\
&&f_{S^h} f_S g^{p0}_{S^h_1S_1\rho} e^{-2\bar{u}_0\Lambda_{S^h}/T-2u_0\Lambda_S/T}\nonumber\\
&&=m_\rho^2
 \biggl\{{1\over 2}f_\rho m_\rho^3\widetilde\Psi^{\{-3\}}-{1\over 2}f_\rho^T m_\rho^2{\cal T}^{\{-2\}}
+f_\rho m_\rho[6\widetilde\Phi^{[-1]}(u_0)+2\widetilde\Psi^{[-1]}(u_0)+{\cal A}^{[-1]}(u_0)]\nonumber\\
&&\phantom{=}+f_\rho^T[{\cal T}^{[0]}(u_0)+2{\cal T}_1^{[0]}(u_0)+2{\cal T}_2^{[0]}(u_0)]Tf_0(\frac{\omega'_c}{T})\biggr\}\,,\nonumber
\\
&&f_{S^h} f_S g^{p1}_{S^h_1S_1\rho} e^{-2\bar{u}_0\Lambda_{S^h}/T-2u_0\Lambda_S/T}\nonumber\\
&&=-m_\rho
 \biggl\{f_\rho m_\rho^2[{\cal V}^{[-1]}(u_0)+{\cal A}^{[-1]}(u_0)]
-f_\rho^T m_\rho[{\cal T}_1^{[0]}(u_0)-{\cal T}_2^{[0]}(u_0)+S^{[0]}(u_0)]Tf_0(\frac{\omega'_c}{T})\nonumber\\
&&\phantom{=}-{1\over 2}f_\rho[{\cal V}^{[1]}(u_0)+{\cal A}^{[1]}(u_0)]T^2f_1(\frac{\omega'_c}{T})\biggr\}\,,\nonumber
\\
&&f_{T^h} f_H g^{s1}_{T^h_1H_1\rho} e^{-2\bar{u}_0\Lambda_{T^h}/T-2u_0\Lambda_H/T}\nonumber\\
&&=-\frac{1}{12\sqrt{2}} m_\rho
\biggl\{f_\rho m_\rho^4[2\Phi+2\widetilde\Phi+2\widetilde\Psi+{\cal A}]^{\{-2\}}\nonumber\\
&&\phantom{=}-4f_\rho^T m_\rho^3[{\cal T}^{[-1]}(u_0)-2{\cal T}_1^{[-1]}(u_0)+2{\cal T}_2^{[-1]}(u_0)-2{\cal T}_4^{[-1]}(u_0)-2\widetilde S^{[-1]}(u_0)]\nonumber\\
&&\phantom{=}-2f_\rho m_\rho^2[2{\cal V}^{[0]}(u_0)+2\Phi^{[0]}(u_0)+2\widetilde\Phi^{[0]}(u_0)+2\widetilde\Psi^{[0]}(u_0)-3{\cal A}^{[0]}(u_0)]Tf_0(\frac{\omega'_c}{T})\nonumber\\
&&\phantom{=}+f_\rho^T m_\rho[{\cal T}^{[1]}(u_0)+4{\cal T}_1^{[1]}(u_0)+2{\cal T}_2^{[1]}(u_0)-2{\cal T}_4^{[1]}(u_0)+4\widetilde S^{[1]}(u_0)]T^2f_1(\frac{\omega'_c}{T})
-f_\rho[{\cal V}^{[2]}(u_0)-2{\cal A}^{[2]}(u_0)]T^3f_2(\frac{\omega'_c}{T})\biggl\}\,,\nonumber
\\
&&f_{T^h} f_H g^{d1}_{T^h_1H_1\rho} e^{-2\bar{u}_0\Lambda_{T^h}/T-2u_0\Lambda_H/T}\nonumber\\
&&=-\frac{1}{4\sqrt{2}} m_\rho
\biggl\{f_\rho m_\rho^2[2\Phi+2\widetilde\Phi+2\widetilde\Psi+{\cal A}]^{\{-2\}}\nonumber\\
&&\phantom{=}-4f_\rho^T m_\rho[{\cal T}^{[-1]}(u_0)+2{\cal T}_1^{[-1]}(u_0)+2{\cal T}_2^{[-1]}(u_0)
+{\cal T}_4^{[-1]}(u_0)-2\widetilde S^{[-1]}(u_0)]
+2f_\rho [{\cal V}^{[0]}(u_0)-2{\cal A}^{[0]}(u_0)]Tf_0(\frac{\omega'_c}{T})\biggl\}\,,\nonumber
\\
&&f_{T^h} f_H g^{d2}_{T^h_1H_1\rho} e^{-2\bar{u}_0\Lambda_{T^h}/T-2u_0\Lambda_H/T}\nonumber\\
&&=\frac{3}{2\sqrt{2}} m_\rho
\biggl\{2f_\rho^Tm_\rho[{\cal T}_1^{[-1]}(u_0)+{\cal T}_4^{[-1]}(u_0)]-f_\rho {\cal V}^{[0]}(u_0)Tf_0(\frac{\omega'_c}{T})\biggl\}\,,\nonumber
\\
&&f_{T^h} f_S g^{p1}_{T^h_1S_1\rho} e^{-2\bar{u}_0\Lambda_{T^h}/T-2u_0\Lambda_S/T}\nonumber\\
&&=-\frac{1}{4\sqrt{2}} m_\rho\biggl\{2f_\rho m_\rho^2[{\cal A}^{[-1]}(u_0)-2{\cal V}^{[-1]}(u_0)]\nonumber\\
&&\phantom{=}+2f_\rho^T m_\rho[{\cal T}_2^{[0]}(u_0)-{\cal T}_1^{[0]}(u_0)+2S^{[0]}(u_0)]Tf_0(\frac{\omega'_c}{T})
-f_\rho[{\cal A}_2^{[1]}(u_0)-2{\cal V}^{[1]}(u_0)]T^2f_1(\frac{\omega'_c}{T})\biggl\}\,,\nonumber
\\
&&f_{T^h} f_S g^{p2}_{T^h_1S_1\rho} e^{-2\bar{u}_0\Lambda_{T^h}/T-2u_0\Lambda_S/T}\nonumber\\
&&=\frac{3}{40\sqrt{2}} m_\rho\biggl\{4f_\rho m_\rho^4\widetilde\Psi^{\{-3\}}
+2f_\rho^T m_\rho^3{\cal T}^{\{-2\}}-4f_\rho m_\rho^2[{\cal A}^{[-1]}(u_0)-4\widetilde\Psi^{[-1]}(u_0)]\nonumber\\
&&\phantom{=}-4f_\rho^Tm_\rho[{\cal T}^{[0]}(u_0)+5{\cal T}_1^{[0]}(u_0)+5{\cal T}_2^{[0]}(u_0)]Tf_0(\frac{\omega'_c}{T})
+6f_\rho{\cal A}^{[1]}(u_0)T^2f_1(\frac{\omega'_c}{T})\biggl\}\,,\nonumber
\\
&&f_{T^h} f_S g^{f2}_{T^h_1S_1\rho} e^{-2\bar{u}_0\Lambda_{T^h}/T-2u_0\Lambda_S/T}\nonumber\\
&&=\frac{3}{4\sqrt{2}} m_\rho
\biggl\{2f_\rho m_\rho^2\widetilde\Psi^{\{-3\}}-f_\rho^T m_\rho{\cal T}^{\{-2\}}
+8f_\rho {\cal A}^{[-1]}(u_0)\biggl\}\,.
\end{eqnarray}

\section{Numerical analysis}\label{numerics}

The parameters in the distribution amplitudes of the $\rho$ meson take their values from \cite{rholcda}.
In this work, we take the values with $\mu=1$ GeV,
realizing that the heavy quark behaves almost as a spectator of the
decay processes in our discussion at the leading order of HQET:
\begin{center}
\setlength\extrarowheight{8pt}
\begin{tabular}{cccccccccccc}
  \hline
  $f^\parallel_\rho$[MeV]&$f^\perp_\rho$[MeV]&$a^\parallel_2$&$a^\perp_2$&$\zeta^\parallel_{3\rho}$
  &$\tilde{\omega}^\parallel_{3\rho}$&$\omega^\parallel_{3\rho}$&$\omega^\perp_{3\rho}$&$\zeta^\parallel_4$
  &$\tilde{\omega}^\parallel_4$&$\zeta^\perp_4$&$\tilde{\zeta}^\perp_4$\\[1.2mm]\hline
  $216(3)$&$165(9)$&$0.15(7)$&$0.14(6)$&$0.030(10)$&$-0.09(3)$&$0.15(5)$&$0.55(25)$
  &$0.07(3)$&$-0.03(1)$&$-0.03(5)$&$-0.08(5)$  \\[1.2mm]
  \hline\\
\end{tabular}
\end{center}

For the mass sum rules of $H$ and $S$,
the working region of the Borel parameter $T$
is about $0.8<T<1.1\ \text{GeV}$ \cite{HSparameter},
which is in the vicinity of that of the mass sum rules for $D^h~(D=H/S/M/T)$~\cite{HeavyHybridparameter}.
So we choose $u_0=1/2$ in our calculation.
The continuum contribution can be subtracted cleanly with this choice.
Asymmetric choice of $u_0$, on the other hand, would
result in a fuzzy continuum substraction \cite{asymmetricpoint}.

The binding energy and the overlapping amplitudes
of doublets $H/S$  \cite{HSparameter} and $H^h/M^h$, $S^h/T^h$ \cite{HeavyHybridparameter}
involved in our numerical analysis
are as follows.

\begin{table}[!h]
\belowrulesep=0pt 
\aboverulesep=0pt 
\renewcommand{\arraystretch}{1.3} 
\doublerulesep 2.2pt
\label{Lambdaf}
\begin{tabular*}{0.8\textwidth}{@{\hspace*{14pt}}@{\extracolsep{\fill}}ccccc@{\hspace*{14pt}}}
\hline\ \\[-4mm]
                     &  $H$              &      $S$         &      $H^h/M^h$      &    $S^h/T^h$\\\hline\ \\[-4.2mm]
$\Lambda$~[GeV]      &  0.50             &     1.15         &      2.0            &    2.5  \\
$f$                  & 0.25~GeV$^{3/2}$  & 0.40~GeV$^{3/2}$ &    1.1~GeV$^{7/2}$  &  1.6~GeV$^{7/2}$ \\[0.5mm]
\hline
\end{tabular*}
\end{table}

\begin{figure}[!h]
\centering
\subfloat[]{\includegraphics[width=3in]{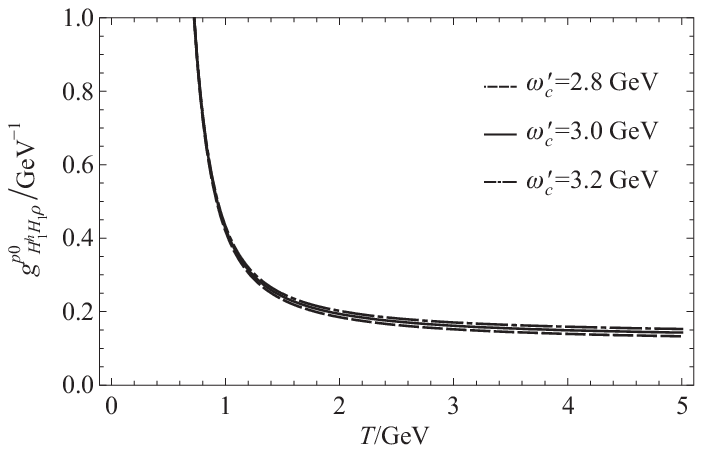}}~~~~~~~~~
\subfloat[]{\includegraphics[width=3in]{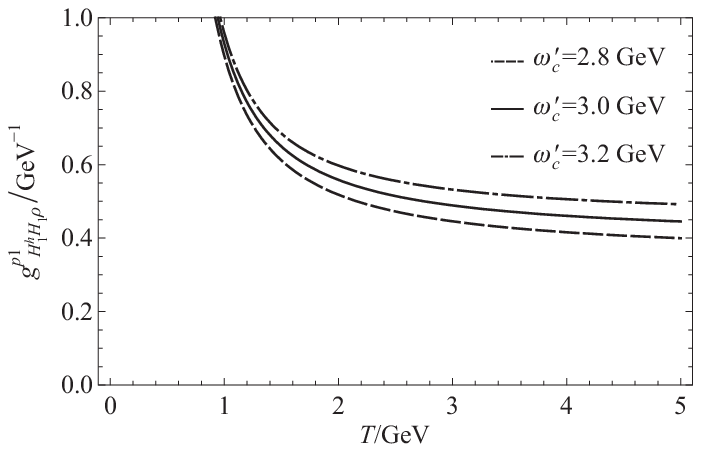}}
\caption{The sum rules for (a) $g_{H^h_1H_1\rho}^{p0}$ and (b) $g_{H^h_1H_1\rho}^{p1}$ with continuum threshold $\omega'_c=2.8, 3.0, 3.2\ \text{GeV}$.}
\label{fig:ghH1H1rho}
\end{figure}

The working region of $T$ is determined by
the insensitivity of the coupling constant to the variation of $T$
and by the requirement that the pole contribution should be not less than 40\%,
We display the sum rules for these $\rho$ couplings with $\omega'_c=2.8, 3.0, 3.2\
\text{GeV}$ in Fig. \ref{fig:ghH1H1rho}.

The following relations arise naturally in our calculation
\begin{eqnarray}\label{proprelation}
&&g^{p0}_{H^hH\rho}\equiv g^{p0}_{H^h_1H_1\rho}=-g^{p0}_{H^h_0H_0\rho}\,,\nonumber\\
&&g^{p1}_{H^hH\rho}\equiv g^{p1}_{H^h_1H_1\rho}=g^{p1}_{H^h_1H_0\rho}=g^{p1}_{H^h_0H_1\rho}\,,\nonumber\\
&&g^{s1}_{H^hS\rho}\equiv g^{s1}_{H^h_1S_1\rho}=g^{s1}_{H^h_0S_1\rho}=g^{s1}_{H^h_1S_0\rho}\,,\nonumber\\
&&g^{d1}_{H^hS\rho}\equiv g^{d1}_{H^h_1S_1\rho}=g^{d1}_{H^h_0S_1\rho}=g^{d1}_{H^h_1S_0\rho}\,,\nonumber\\
&&g^{p1}_{M^hH\rho}\equiv g^{p1}_{M^h_1H_1\rho}=-{1\over 2}g^{p1}_{M^h_1H_0\rho}={\sqrt{6}\over 3}g^{p1}_{M^h_2H_1\rho}\,,\nonumber\\
&&g^{p2}_{M^hH\rho}\equiv g^{p2}_{M^h_1H_1\rho}={\sqrt{6}\over 2}g^{p2}_{M^h_2H_0\rho}=-\sqrt{6}g^{p2}_{M^h_2H_1\rho}\,,\nonumber\\
&&g^{f2}_{M^hH\rho}\equiv g^{f2}_{M^h_1H_1\rho}={\sqrt{6}\over 2}g^{f2}_{M^h_2H_0\rho}=-\sqrt{6}g^{f2}_{M^h_2H_1\rho}\,,\nonumber\\
&&g^{s1}_{M^hS\rho}\equiv g^{s1}_{M^h_1S_1\rho}=-{1\over 2}g^{s1}_{M^h_1S_0\rho}={\sqrt{6}\over 3}g^{s1}_{M^h_2S_1\rho}\,,\nonumber\\
&&g^{d1}_{M^hS\rho}\equiv g^{d1}_{M^h_1S_1\rho}={1\over 2}g^{d1}_{M^h_1S_0\rho}=-{\sqrt{6}\over 3}g^{d1}_{M^h_2S_1\rho}\,,\nonumber\\
&&g^{d2}_{M^hS\rho}\equiv g^{d2}_{M^h_1S_1\rho}={\sqrt{6}\over 2}g^{d2}_{M^h_2S_0\rho}=\sqrt{6}g^{d2}_{M^h_2S_1\rho}\,,\nonumber\\
&&g^{s1}_{S^hH\rho}\equiv g^{s1}_{S^h_1H_1\rho}=-g^{s1}_{S^h_0H_1\rho}=g^{s1}_{S^h_1H_0\rho}\,,\nonumber\\
&&g^{d1}_{S^hH\rho}\equiv g^{d1}_{S^h_1H_1\rho}=-g^{d1}_{S^h_0H_1\rho}=g^{d1}_{S^h_1H_0\rho}\,,\nonumber\\
&&g^{p0}_{S^hS\rho}\equiv g^{p0}_{S^h_1S_1\rho}=g^{p0}_{S^h_0S_0\rho}\,,\nonumber\\
&&g^{p1}_{S^hS\rho}\equiv g^{p1}_{S^h_1S_1\rho}=g^{p1}_{S^h_1S_0\rho}=-g^{p1}_{S^h_0S_1\rho}\,,\nonumber\\
&&g^{s1}_{T^hH\rho}\equiv g^{s1}_{T^h_1H_1\rho}=-{1\over 2}g^{s1}_{T^h_1H_0\rho}=-{\sqrt{6}\over 3}g^{s1}_{T^h_2H_1\rho}\,,\nonumber\\
&&g^{d1}_{T^hH\rho}\equiv g^{d1}_{T^h_1H_1\rho}=-{1\over 2}g^{d1}_{T^h_1H_0\rho}=-{\sqrt{6}\over 3}g^{d1}_{T^h_2H_1\rho}\,,\nonumber\\
&&g^{d2}_{T^hH\rho}\equiv g^{d2}_{T^h_1H_1\rho}=-{\sqrt{6}\over 2}g^{d2}_{T^h_2H_0\rho}=-\sqrt{6}g^{d2}_{T^h_2H_1\rho}\,,\nonumber\\
&&g^{p1}_{T^hS\rho}\equiv g^{p1}_{T^h_1S_1\rho}=-{1\over 2}g^{p1}_{T^h_1S_0\rho}=-{\sqrt{6}\over 3}g^{p1}_{T^h_2S_1\rho}\,,\nonumber\\
&&g^{p2}_{T^hS\rho}\equiv g^{p2}_{T^h_1S_1\rho}=-{\sqrt{6}\over 2}g^{p2}_{T^h_2S_0\rho}=\sqrt{6}g^{p2}_{T^h_2S_1\rho}\,,\nonumber\\
&&g^{f2}_{T^hS\rho}\equiv g^{f2}_{T^h_1S_1\rho}=-{\sqrt{6}\over 2}g^{f2}_{T^h_2S_0\rho}=\sqrt{6}g^{f2}_{T^h_2S_1\rho}\,.
\end{eqnarray}
These simple proportional relations among the obtained couplings
result from the heavy quark flavor-spin symmetry.
They also justify our construction of the interpolating currents for heavy hybrid mesons.
The spin of the interpolating currents can be deduced from the symmetry of their Lorentz indices.
The $P$ parity can be obtained directly from the $P$-transformation property of these currents.
The tensor structure of the correlation functions considered above verifies their $J^P$ quantum numbers.
For example, if the $J^P$ quantum number of
$
J^{\dag\alpha}_{T^h_1}
=\bar{h}_vg_s\gamma_5[3G_t^{\alpha\beta}\gamma_\beta+i\gamma^\alpha_t\sigma_t\cdot G] q
$
is not $1^+$, the tensor structure of the correlation function
\begin{eqnarray}
i\int dx\ e^{-ik\cdot x}\langle\rho(q)|J_{H_1}^\beta(0)J^{\dag\alpha}_{T^h_1}(x)|0\rangle
\end{eqnarray}
cannot include (only) $s1$, $d1$ and $d2$.

Furthermore, if
$
J^{\dag\alpha}_{T^h_1}=J^{\dag\alpha}_{T^h_{10}}+\lambda J^{\dag\alpha}_{S^h_{10}}~(\lambda\neq 0),
$
where
$J^{\dag\alpha}_{T^h_{10}}$ and $J^{\dag\alpha}_{S^h_{10}}$ are pure interpolating currents with $j_l=3/2$ and $j_l=1/2$, respectively, we have
\begin{eqnarray*}
i\int dx\ e^{-ik\cdot x}\langle\rho(q)|J_{H_1}^\beta(0)J^{\dag\alpha}_{T^h_{10}}(x)|0\rangle
&=&I\epsilon^{\alpha\beta e^*v}G^{s1}_{T^h_{10}H_1\rho}
+I\Big[\epsilon^{\alpha\beta qv}(e^*\cdot q_t)-\frac{1}{3}\epsilon^{\alpha\beta e^*v}q_t^2\Big]G^{d1}_{T^h_{10}H_1\rho}\nonumber\\
&&+\,I\Big[\epsilon^{\alpha e q v}q_t^\beta+\epsilon^{\beta e q v}q_t^\alpha)\Big] G^{d2}_{T^h_{10}H_1\rho}\,,\\
i\int dx\ e^{-ik\cdot x}\langle\rho(q)|J_{H_1}^\beta(0)J^{\dag\alpha}_{S^h_{10}}(x)|0\rangle
&=&I\epsilon^{\alpha\beta e^*v}G^{s1}_{S^h_{10}H_1\rho}
+I\Big[\epsilon^{\alpha\beta qv}(e^*\cdot q_t)-\frac{1}{3}\epsilon^{\alpha\beta e^*v}q_t^2\Big]G^{d1}_{S^h_{10}H_1\rho},
\end{eqnarray*}
and
\begin{eqnarray*}
i\int dx\ e^{-ik\cdot x}\langle\rho(q)|J_{H_0}(0)J^{\dag\alpha}_{T^h_{10}}(x)|0\rangle
&=&Ie^{*\alpha}_tG^{s1}_{T^h_{10}H_0\rho}+I
\Big[q_t^\alpha(e^*\cdot q_t)-\frac{1}{3}e^{*\alpha}_tq_t^2\Big]G^{d1}_{T^h_{10}H_0\rho},
\\
i\int dx\ e^{-ik\cdot x}\langle\rho(q)|J_{H_0}(0)J^{\dag\alpha}_{S^h_{10}}(x)|0\rangle
&=&Ie^{*\alpha}_tG^{s1}_{S^h_{10}H_0\rho}+I
\Big[q_t^\alpha(e^*\cdot q_t)-\frac{1}{3}e^{*\alpha}_tq_t^2\Big]G^{d1}_{S^h_{10}H_0\rho}.
\end{eqnarray*}
Now let us focus on the $s1$ part. It is straightforward that
$$
G^{s1}_{T^h_{10}H_1\rho}=c_1G^{s1}_{T^h_{10}H_0\rho},~~~G^{s1}_{S^h_{10}H_1\rho}=c_2G^{s1}_{S^h_{10}H_0\rho},
$$
due to the heavy quark flavor-spin symmetry,
therefore
\begin{eqnarray*}
&&G^{s1}_{T^h_{1}H_1\rho}=c_1G^{s1}_{T^h_{10}H_0\rho}+\lambda c_2G^{s1}_{S^h_{10}H_0\rho},\\
&&G^{s1}_{T^h_{1}H_0\rho}=G^{s1}_{T^h_{10}H_0\rho}+\lambda G^{s1}_{S^h_{10}H_0\rho},\\
&&c_1
=G^{s1}_{T^h_{10}H_1\rho}/G^{s1}_{T^h_{10}H_0\rho}
=G^{s1}_{T^h_{1}H_1\rho}/G^{s1}_{T^h_{1}H_0\rho}
=g^{s1}_{T^h_{1}H_1\rho}/g^{s1}_{T^h_{1}H_0\rho},\\
&&c_2
=G^{s1}_{S^h_{10}H_1\rho}/G^{s1}_{S^h_{10}H_0\rho}
=G^{s1}_{S^h_{1}H_1\rho}/G^{s1}_{S^h_{1}H_0\rho}
=g^{s1}_{S^h_{1}H_1\rho}/g^{s1}_{S^h_{1}H_0\rho}.
\end{eqnarray*}
When $G^{s1}_{T^h_{1}H_1\rho}$ is proportional to $G^{s1}_{T^h_{1}H_0\rho}$,
we have $c_1=c_2$, namely,
$$
g^{s1}_{T^h_{1}H_1\rho}/g^{s1}_{T^h_{1}H_0\rho}= g^{s1}_{S^h_{1}H_1\rho}/g^{s1}_{S^h_{1}H_0\rho}.
$$
This is inconsistent with the results (see Eq.~(\ref{proprelation})) we just obtained:
$
g^{s1}_{T^h_{1}H_1\rho}/g^{s1}_{T^h_{1}H_0\rho}\neq g^{s1}_{S^h_{1}H_1\rho}/g^{s1}_{S^h_{1}H_0\rho}.
$
This implies $J^{\dag\alpha}_{T^h_1}=J^{\dag\alpha}_{T^h_{10}}$.
In other words, the interpolating current $J^{\dag\alpha}_{T^h_1}$
carries $j_l=3/2$.
The $J$, $P$ and $j_l$ quantum numbers of other interpolating currents can be verified in a similar way.

\begin{table}[htb]
\belowrulesep=12pt 
\aboverulesep=2pt 
\renewcommand{\arraystretch}{1.3} 
\doublerulesep 2.2pt
\begin{tabular*}{.88\textwidth}{@{\hspace*{6pt}}@{\extracolsep{\fill}}cccccccccc@{\hspace*{6pt}}}
\hline\ \\[-3.8mm]
  $g_{H^h_1H_1\rho}^{p0}$  &   $g_{H^h_1H_1\rho}^{p1}$  &  $g_{H^h_1S_1\rho}^{s1}$   &   $g_{H^h_1S_1\rho}^{d1}$  &   $g_{M^h_1H_1\rho}^{p1}$  &  $g_{M^h_1H_1\rho}^{p2}$   &
  $g_{M^h_1H_1\rho}^{f2}$  &   $g_{M^h_1S_1\rho}^{s1}$  &  $g_{M^h_1S_1\rho}^{d1}$   &
  $g_{M^h_1S_1\rho}^{d2}$     \\[2mm]
  0.2&0.5&  $-1.0$ &  0.09   &  0.12  &     0.21  &   $-0.12$   &   $-0.01$  &  $0.02$  &  0.07\\[.5mm]\hline\ \\[-3.8mm]
  $g_{S^h_1H_1\rho}^{s1}$  &  $g_{S^h_1H_1\rho}^{d1}$   &
  $g_{S^h_1S_1\rho}^{p0}$  &   $g_{S^h_1S_1\rho}^{p1}$  &  $g_{T^h_1H_1\rho}^{s1}$   &  $g_{T^h_1H_1\rho}^{d1}$  &   $g_{T^h_1H_1\rho}^{d2}$  &  $g_{T^h_1S_1\rho}^{p1}$   &  $g_{T^h_1S_1\rho}^{p2}$  &   $g_{T^h_1S_1\rho}^{f2}$  \\[2mm]
  1.4  & $-0.4$  & 0.14  & $-0.22$ & $-0.18$& 0.09 & $-0.09$ & $-0.02$ &  $-0.15$  &   0.06 \\[.5mm]
\hline
\end{tabular*}
\caption{The absolute values of the coupling constants.
The units of the $P$-, $D$-wave and $F$-wave coupling constants are $\text{GeV}^{-1}$,
$\text{GeV}^{-2}$ and $\text{GeV}^{-3}$ respectively.
}
\label{tablecc}
\end{table}

The final values of these couplings are listed in Table \ref{tablecc}.
In most channels they are rather small, which may be attributed to
the fading of the gluon degree of freedom in the decay.

\section{conclusion}\label{conclusion}

At the heavy quark limit,
we have constructed interpolating currents respecting the flavor-spin symmetry for $q\bar{Q}g$ and $q\bar{Q}$.
With these currents, the $\rho$ meson couplings
between $q\bar{Q}g$ and $q\bar{Q}$ have been worked out by means of LCQSR.
The derived sum rules rely mildly on the Borel parameters
in their working regions.
The resulting coupling constants are rather small in most cases.

The main error of our calculation
originate from the inaccuracy of LCQSR:
truncation of the OPE near the light-cone,
the uncertainty of the parameters in the light-cone wave functions,
the dependence of the coupling constant
on the continuum threshold $\omega_c$ and the
Borel parameter in the working region,
the uncertainty of the binding energy $\bar{\Lambda}$'s
and the overlapping amplitudes $f$'s.
As far as the charm quark is concerned, the $1/m_Q$ correction may be significant,
while the correction from the finite mass of the bottom quark should be negligible.

We hope that our calculation may be helpful to experimental searches for these heavy hybrid mesons
and the understanding of their strong interaction with conventional heavy mesons.
Moreover, the coupling constants calculated in our work might
shed further light on the nature of the \textit{XYZ} mesons.

\section*{Acknowledgments}

This work is supported by the National Natural Science Foundation of
China under Grants No. 11105007.


\appendix

\section{The $\rho$ decay amplitudes of heavy hybrid mesons}\label{appendixrho}

The $\rho$ decay amplitudes considered in the text are as follows.
\begin{eqnarray}
\mathcal {M}(H^h_0\rightarrow H_0+\rho)
&=&I(e^*\cdot q_t)g^{p0}_{H^h_0H_0\rho},
\\
\mathcal {M}(H^h_0\rightarrow H_1+\rho)
&=&I\epsilon^{\epsilon^* e^* q v}g^{p1}_{H^h_0H_1\rho},
\\
\mathcal {M}(H^h_1\rightarrow H_0+\rho)
&=&I\epsilon^{\eta e^* q v}g^{p1}_{H^h_1H_0\rho},
\\
\mathcal {M}(H^h_1\rightarrow H_1+\rho)
&=&I(e^*\cdot q_t)(\epsilon^*\cdot \eta_t)g^{p0}_{H^h_1H_1\rho}+I
\Big[(e^*\cdot \eta_t)(\epsilon^*\cdot q_t)-(e^*\cdot \epsilon^*_t)(\eta\cdot q_t)\Big]g^{p1}_{H^h_1H_1\rho},
\\
\mathcal {M}(H^h_0\rightarrow S_1+\rho)
&=&I(e^*\cdot \epsilon^*_t)g^{s1}_{H^h_0S_1\rho}+I
\Big[(\epsilon^*\cdot q_t)(e^*\cdot q_t)-\frac{1}{3}(e^*\cdot \epsilon^*_t)q_t^2\Big]g^{d1}_{H^h_0S_1\rho},
\\
\mathcal {M}(H^h_1\rightarrow S_0+\rho)
&=&I(e^*\cdot \eta_t)g^{s1}_{H^h_1S_0\rho}+I
\Big[(\eta\cdot q_t)(e^*\cdot q_t)-\frac{1}{3}(e^*\cdot \eta_t)q_t^2\Big]g^{d1}_{H^h_1S_0\rho},
\\
\mathcal {M}(H^h_1\rightarrow S_1+\rho)
&=&I\epsilon^{\eta\epsilon^*e^*v}g^{s1}_{H^h_1S_1\rho}+I
\Big[\epsilon^{\eta\epsilon^*qv}(e^*\cdot q_t)-\frac{1}{3}\epsilon^{\eta\epsilon^*e^*v}q_t^2\Big]g^{d1}_{H^h_1S_1\rho},
\\
\mathcal {M}(M^h_1\rightarrow H_0+\rho)
&=&I\epsilon^{\eta e^*qv}g^{p1}_{M^h_1H_0\rho}\,,
\\
\mathcal {M}(M^h_1\rightarrow H_1+\rho)
&=&Ii\left[(\eta\cdot e_t^*)(\epsilon^*\cdot q_t)-(\eta\cdot q_t)(\epsilon^*\cdot e^*_t)\right]g^{p1}_{M^h_1H_1\rho}\nonumber\\
&&+Ii\left[(\eta\cdot e_t^*)(\epsilon^*\cdot q_t)+(\eta\cdot q_t)(\epsilon^*\cdot e^*_t)-\frac{2}{3}(\eta\cdot \epsilon^*_t)(e^*\cdot q_t)\right]g^{p2}_{M^h_1H_1\rho}\nonumber\\
&&+Ii\biggl\lbrace(\eta\cdot q_t)(\epsilon^*\cdot q_t)(e^*\cdot q_t)
-\frac{q_t^2}{5}\left[(\eta\cdot \epsilon^*_t)(e^*\cdot q_t)
+(\eta\cdot e_t^*)(\epsilon^*\cdot q_t)
+(\eta\cdot q_t)(\epsilon^*\cdot e^*_t)\right]\bigg\rbrace g^{f2}_{M^h_1H_1\rho}\,,\nonumber
\\\ \\
\mathcal {M}(M^h_2\rightarrow H_0+\rho)
&=&2Ii\eta_{\alpha_1\alpha_2}\left[e_t^{*\alpha_1}q_t^{\alpha_2}-\frac{1}{3}g_t^{\alpha_1\alpha_2}(e^*\cdot q_t)\right]g^{p2}_{M^h_2H_0\rho}\nonumber\\
&&+Ii\eta_{\alpha_1\alpha_2}\biggl\lbrace q_t^{\alpha_1}q_t^{\alpha_2}(e^*\cdot q_t)-\frac{q_t^2}{5}
\left[g_t^{\alpha_1\alpha_2}(e^*\cdot q_t)+2e_t^{*\alpha_1}q_t^{\alpha_2}\right]\biggl\rbrace g^{f2}_{M^h_2H_0\rho}\,,
\\
\mathcal {M}(M^h_2\rightarrow H_1+\rho)
&=&
2I\eta_{\alpha_1\alpha_2}\left[-\varepsilon^{\alpha_1 e^*qv}\epsilon^{*\alpha_2}_t
+\frac{1}{3}g_t^{\alpha_1\alpha_2}\varepsilon^{\epsilon^* e^*qv}\right]g^{p1}_{M^h_2H_1\rho}\nonumber\\
&&
+2I\eta_{\alpha_1\alpha_2}\left[\varepsilon^{\alpha_1\epsilon^* e^*v}q_t^{\alpha_2}
+\varepsilon^{\alpha_1\epsilon^* qv}e_t^{*\alpha_2}\right]g^{p2}_{M^h_2H_1\rho}\nonumber\\
&&
+2I\eta_{\alpha_1\alpha_2}\biggl\lbrace
\varepsilon^{\alpha_1\epsilon^* qv}q_t^{\alpha_2}(e^*\cdot q_t)
-\frac{q_t^2}{5}\left[\varepsilon^{\alpha_1\epsilon^* q v}e_t^{*\alpha_2}+\varepsilon^{\alpha_1\epsilon^* e^* v}q_t^{\alpha_2}\right]
\biggl\rbrace g^{f2}_{M^h_2H_1\rho}\,,
\\
\mathcal {M}(M^h_1\rightarrow S_0+\rho)
&=&Ii(\eta\cdot e^*_t)g^{s1}_{M^h_1S_0\rho}+Ii\left[(\eta\cdot q_t)(e^*\cdot q_t)-\frac{1}{3}(\eta\cdot e^*_t)q_t^2\right]g^{d1}_{M^h_1S_0\rho}\,,
\\
\mathcal {M}(M^h_1\rightarrow S_1+\rho)
&=&I\epsilon^{\eta\epsilon^*e^*v}g^{s1}_{M^h_1S_1\rho}+I\left[\epsilon^{\eta\epsilon^*qv}(e^*\cdot q_t)-\frac{1}{3}\epsilon^{\eta\epsilon^*e^*v}q_t^2\right]g^{d1}_{M^h_1S_1\rho}\nonumber\\
&&+I\left[\epsilon^{\eta e^*qv}(\epsilon^*\cdot q_t)+\epsilon^{\epsilon^*e^*qv}(\eta\cdot q_t)\right]g^{d2}_{M^h_1S_1\rho}\,,
\\
\mathcal {M}(M^h_2\rightarrow S_0+\rho)
&=&2I\eta_{\alpha_1\alpha_2}\epsilon^{\alpha_1e^*qv}q_t^{\alpha_2}g^{d2}_{M^h_2S_0\rho}\,,
\\
\mathcal {M}(M^h_2\rightarrow S_1+\rho)
&=&2Ii\eta_{\alpha_1\alpha_2}\left[\epsilon_t^{*\alpha_1}e_t^{*\alpha_2}
-\frac{1}{3}g_t^{\alpha_1\alpha_2}(\epsilon^*\cdot e_t^*)\right] g^{s1}_{M^h_2S_1\rho}\nonumber\\
&&+2Ii\eta_{\alpha_1\alpha_2}\biggl\lbrace\left[\epsilon_t^{*\alpha_1}q_t^{\alpha_2}
-\frac{1}{3}g_t^{\alpha_1\alpha_2}(\epsilon^*\cdot q_t)\right](e^*\cdot q_t)
-\frac{q_t^2}{3}\left[\epsilon_t^{*\alpha_1}e_t^{*\alpha_2}
-\frac{1}{3}g_t^{\alpha_1\alpha_2}(\epsilon^*\cdot e_t^*)\right]\biggl\rbrace g^{d1}_{M^h_2S_1\rho}\nonumber\\
&&+2Ii\eta_{\alpha_1\alpha_2}\biggl\lbrace2\left[e_t^{*\alpha_1}q_t^{\alpha_2}(\epsilon^*\cdot q_t)-q_t^{\alpha_1}q_t^{\alpha_2}(\epsilon^*\cdot e_t^*)\right]\nonumber\\
&&+\left[\epsilon^{*\alpha_1}_tq_t^{\alpha_2}-g_t^{\alpha_1\alpha_2}(\epsilon^*\cdot q_t)\right](e^*\cdot q_t)
-\left[\epsilon^{*\alpha_1}_te_t^{*\alpha_2}-g_t^{\alpha_1\alpha_2}(\epsilon^*\cdot e^*_t)\right]q_t^2\biggl\rbrace g^{d2}_{M^h_2S_1\rho}\,,
\\
\mathcal {M}(S^h_0\rightarrow S_0+\rho)
&=&I(e^*\cdot q_t)g^{p0}_{S^h_0S_0\rho},
\\
\mathcal {M}(S^h_0\rightarrow S_1+\rho)
&=&I\epsilon^{\epsilon^* e^* q v}g^{p1}_{S^h_0S_1\rho},
\\
\mathcal {M}(S^h_1\rightarrow S_0+\rho)
&=&I\epsilon^{\eta e^* q v}g^{p1}_{S^h_1S_0\rho},
\\
\mathcal {M}(S^h_1\rightarrow S_1+\rho)
&=&I(e^*\cdot q_t)(\epsilon^*\cdot \eta_t)g^{p0}_{S^h_1S_1\rho}+I
\Big[(e^*\cdot \eta_t)(\epsilon^*\cdot q_t)-(e^*\cdot \epsilon^*_t)(\eta\cdot q_t)\Big]g^{p1}_{S^h_1S_1\rho},
\\
\mathcal {M}(S^h_0\rightarrow H_1+\rho)
&=&I(e^*\cdot \epsilon^*_t)g^{s1}_{S^h_0H_1\rho}+I
\Big[(\epsilon^*\cdot q_t)(e^*\cdot q_t)-\frac{1}{3}(e^*\cdot \epsilon^*_t)q_t^2\Big]g^{d1}_{S^h_0H_1\rho},
\\
\mathcal {M}(S^h_1\rightarrow H_0+\rho)
&=&I(e^*\cdot \eta_t)g^{s1}_{S^h_1H_0\rho}+I
\Big[(\eta\cdot q_t)(e^*\cdot q_t)-\frac{1}{3}(e^*\cdot \eta_t)q_t^2\Big]g^{d1}_{S^h_1H_0\rho},
\\
\mathcal {M}(S^h_1\rightarrow H_1+\rho)
&=&I\epsilon^{\eta\epsilon^*e^*v}g^{s1}_{S^h_1H_1\rho}+I
\Big[\epsilon^{\eta\epsilon^*qv}(e^*\cdot q_t)-\frac{1}{3}\epsilon^{\eta\epsilon^*e^*v}q_t^2\Big]g^{d1}_{S^h_1H_1\rho},
\\
\mathcal {M}(T^h_1\rightarrow H_0+\rho)&=&I(e^*\cdot \eta_t)g^{s1}_{T^h_1H_0\rho}+I
\Big[(\eta\cdot q_t)(e^*\cdot q_t)-\frac{1}{3}(e^*\cdot \eta_t)q_t^2\Big]g^{d1}_{T^h_1H_0\rho},
\\
\mathcal {M}(T^h_1\rightarrow H_1+\rho)&=&I\epsilon^{\eta\epsilon^*e^*v}g^{s1}_{T^h_1H_1\rho}
+I\Big[\epsilon^{\eta\epsilon^*qv}(e^*\cdot q_t)-\frac{1}{3}\epsilon^{\eta\epsilon^*e^*v}q_t^2\Big]g^{d1}_{T^h_1H_1\rho}\nonumber\\
&&+I\Big[\epsilon^{\eta e q v}(q_t\cdot \epsilon^*)+\epsilon^{\epsilon^* e q v}(q_t\cdot \eta)\Big] g^{d2}_{T^h_1H_1\rho},
\\
\mathcal {M}(T^h_1\rightarrow S_0+\rho)&=&I\epsilon^{\eta e^* q v}g^{p1}_{T^h_1S_0\rho},
\\
\mathcal {M}(T^h_1\rightarrow S_1+\rho)&=&I\Big[(e^*\cdot \eta_t)(\epsilon^*\cdot q_t)-(e^*\cdot \epsilon^*_t)(\eta\cdot q_t)\Big]g^{p1}_{T^h_1S_1\rho}\nonumber\\
&&+I\Big[(e_t^*\cdot\eta) (q_t\cdot \epsilon^*)+(q_t\cdot\eta) (e^*_t\cdot \epsilon^*)-\frac{2}{3}(e^*_t\cdot \eta)(e^*\cdot q_t)\Big]g^{p2}_{T^h_1S_1\rho}\nonumber\\
&&+I\Big\{(q_t\cdot\eta) (q_t\cdot\epsilon^*)(e^*\cdot q_t)\nonumber\\
&&-\frac{q_t^2}{5}\Big[(\eta_t\cdot \epsilon^*)(e^*\cdot q_t)+(e_t^*\cdot\eta) (q_t\cdot \epsilon^*)+(q_t\cdot\eta) (e^*_t\cdot \epsilon^*)\Big]\Big\}g^{f2}_{T^h_1S_1\rho},
\\
\mathcal {M}(T^h_2\rightarrow H_0+\rho)&=&I\eta_{\alpha_1\alpha_2}(\epsilon^{\alpha_1 e^* q v}q_t^{\alpha_2}+\epsilon^{\alpha_2 e^* q v}q_t^{\alpha_1})g^{d2}_{T^h_2H_0\rho},
\\
\mathcal {M}(T^h_2\rightarrow H_1+\rho)&=&I\eta_{\alpha_1\alpha_2}\Big[\epsilon^{*\alpha_1}e_t^{*\alpha_2}+\epsilon^{*\alpha_2}e_t^{*\alpha_1}
-\frac{2}{3}g_t^{\alpha_1\alpha_2}(e^*_t\cdot\epsilon^*)\Big] g^{s1}_{T^h_2H_1\rho}\nonumber\\
&&+I\eta_{\alpha_1\alpha_2}\Big\{\Big[\epsilon^{*\alpha_1}q_t^{\alpha_2}+\epsilon^{*\alpha_2}q_t^{\alpha_1}
-\frac{2}{3}g_t^{\alpha_1\alpha_2}(q_t\cdot\epsilon^*)\Big](e^*\cdot q_t)\nonumber\\
&&-\frac{1}{3}\Big[\epsilon^{*\alpha_1}e_t^{*\alpha_2}+\epsilon^{*\alpha_2}e_t^{*\alpha_1}
-\frac{2}{3}g_t^{\alpha_1\alpha_2}(e^*_t\cdot\epsilon^*)\Big]q_t^2\Big\}g^{d1}_{T^h_2H_1\rho}\nonumber\\
&&+I\eta_{\alpha_1\alpha_2}\Big\{2\Big[e_t^{*\alpha_1}q_t^{\alpha_2}(q_t\cdot\epsilon^*)+q_t^{\alpha_1}e_t^{*\alpha_2}(q_t\cdot\epsilon^*)-2q_t^{\alpha_1}q_t^{\alpha_2}(e^*_t\cdot\epsilon^*)\Big]\nonumber\\
&&+\Big[\epsilon_t^{*\alpha_1}q_t^{\alpha_2}+\epsilon_t^{*\alpha_2}q_t^{\alpha_1}-2g_t^{\alpha_1\alpha_2}(q_t\cdot\epsilon^*)\Big](e^*\cdot q_t)\nonumber\\
&&-\Big[\epsilon_t^{*\alpha_1}e_t^{*\alpha_2}+\epsilon_t^{*\alpha_2}e_t^{*\alpha_1}-2g_t^{\alpha_1\alpha_2}(e^*_t\cdot\epsilon^*)\Big]q_t^2\Big\}g^{d2}_{T^h_2H_1\rho},
\\
\mathcal {M}(T^h_2\rightarrow S_0+\rho)&=&I\eta_{\alpha_1\alpha_2}\Big[e_t^{*\alpha_1}q_t^{\alpha_2}+q_t^{\alpha_1}e_t^{*\alpha_2}
-\frac{2}{3}g_t^{\alpha_1\alpha_2}(e^*\cdot q_t)\Big]g^{p2}_{T^h_2S_0\rho}\nonumber\\
&&+I\eta_{\alpha_1\alpha_2}\Big\{q_t^{\alpha_1}q_t^{\alpha_2}(e^*\cdot q_t)-\frac{q_t^2}{5}
\Big[g_t^{\alpha_1\alpha_2}(e^*\cdot q_t)+e_t^{*\alpha_1} q_t^{\alpha_2}+q_t^{\alpha_1} e_t^{*\alpha_2}\Big]\Big\}g^{f2}_{T^h_2S_0\rho},
\\
\mathcal {M}(T^h_2\rightarrow S_1+\rho)
&=&I\eta_{\alpha_1\alpha_2}\Big[-\epsilon^{\alpha_1 e^* q v}\epsilon^{*\alpha_2}_t-\epsilon^{\alpha_2 e^* q v}\epsilon^{*\alpha_1}_t
+\frac{2}{3}g_t^{\alpha_1\alpha_2}\epsilon^{\epsilon^* e^* q v}\Big]g^{p1}_{T^h_2S_1\rho}\nonumber\\
&&+I\eta_{\alpha_1\alpha_2}\Big[\epsilon^{\alpha_1\epsilon^* e^* v}q_t^{\alpha_2}+\epsilon^{\alpha_2\epsilon^* e^* v}q_t^{\alpha_1}
+\epsilon^{\alpha_1\epsilon^* q v}e_t^{\alpha_2}+\epsilon^{\alpha_2\epsilon^* q v}e_t^{*\alpha_1}\Big]g^{p2}_{T^h_2S_1\rho}\nonumber\\
&&+I\eta_{\alpha_1\alpha_2}\Big\{\epsilon^{\alpha_1\epsilon^* q v}q_t^{\alpha_2}(e^*\cdot q_t)
+\epsilon^{\alpha_2\epsilon^* q v}q_t^{\alpha_1}(e^*\cdot q_t)\nonumber\\
&&-\frac{q_t^2}{5}\Big[\epsilon^{\alpha_1\epsilon^* q v}e_t^{*\alpha_2}+\epsilon^{\alpha_2\epsilon^* q v}e_t^{*\alpha_1}
+\epsilon^{\alpha_1\epsilon^* e^* v}q_t^{\alpha_2}+\epsilon^{\alpha_2\epsilon^* e^* v}q_t^{\alpha_1}\Big]\Big\}g^{f2}_{T^h_2S_1\rho}.
\end{eqnarray}

\section{The definitions of the $\rho$ meson light-cone distribution amplitudes}\label{appendixLCDA}

The definitions of the distribution amplitudes of the $\rho$ meson
used in the text read as \cite{rholcda,rhoparameter}
\begin{eqnarray}
\langle 0|\bar u(z) \gamma_{\mu} d(-z)|\rho^-(P,\lambda)\rangle &=&
f_{\rho} m_{\rho} \left[ p_{\mu} \frac{e^{(\lambda)}\cdot z}{p \cdot
z} \int_{0}^{1} \!du\, e^{i \xi p \cdot z} \varphi_{\parallel}(u,
\mu^{2}) \right.
+ e^{(\lambda)}_{\perp \mu}
\int_{0}^{1} \!du\, e^{i \xi p \cdot z} g_{\perp}^{(v)}(u, \mu^{2})
\nonumber \\
& & \left.- \frac{1}{2}z_{\mu}
\frac{e^{(\lambda)}\cdot z }{(p \cdot z)^{2}} m_{\rho}^{2}
\int_{0}^{1} \!du\, e^{i \xi p \cdot z} g_{3}(u, \mu^{2}) \right]\,,\nonumber \\
\langle 0|\bar u(z) \gamma_{\mu} \gamma_{5}
d(-z)|\rho^-(P,\lambda)\rangle &=&
 \frac{1}{2}f_{\rho}
m_{\rho} \varepsilon_{\mu}^{\phantom{\mu}\nu \alpha \beta}
e^{(\lambda)}_{\perp \nu} p_{\alpha} z_{\beta}
\int_{0}^{1} \!du\, e^{i \xi p \cdot z} g^{(a)}_{\perp}(u, \mu^{2})\,,\nonumber
\\
\langle 0|\bar u(z) \sigma_{\mu \nu} d(-z)|\rho^-(P,\lambda)\rangle
&=& i f_{\rho}^{T} \left[ ( e^{(\lambda)}_{\perp \mu}p_\nu -
e^{(\lambda)}_{\perp \nu}p_\mu ) \int_{0}^{1} \!du\, e^{i \xi p
\cdot z} \varphi_{\perp}(u, \mu^{2}) \right.
\nonumber \\
& &+ (p_\mu z_\nu - p_\nu z_\mu )
\frac{e^{(\lambda)} \cdot z}{(p \cdot z)^{2}}
m_{\rho}^{2}
\int_{0}^{1} \!du\, e^{i \xi p \cdot z} h_\parallel^{(t)} (u, \mu^{2})
\nonumber \\
& & \left.+ \frac{1}{2}
(e^{(\lambda)}_{\perp \mu} z_\nu -e^{(\lambda)}_{\perp \nu} z_\mu)
\frac{m_{\rho}^{2}}{p \cdot z}
\int_{0}^{1} \!du\, e^{i \xi p \cdot z} h_{3}(u, \mu^{2}) \right]\,,\nonumber \\
\langle 0|\bar u(z)d(-z)|\rho^-(P,\lambda)\rangle
&=&-if_{\rho}^{T}(e^{(\lambda)}z)m_\rho^2 \int_{0}^{1} \!du\, e^{i \xi p \cdot z} h_\parallel^{(s)} (u, \mu^{2})\,.
\end{eqnarray}
The distribution amplitude $\varphi_\parallel$ and $\varphi_\perp$
are of twist-2, $g_\perp^{(v)}$, $g_\perp^{(a)}$,
$h_\parallel^{(s)}$ and $h_\parallel^{(t)}$ are twist-3 and $g_3$,
$h_3$ are twist-4. All functions
$\phi=\{\varphi_\parallel,\varphi_\perp,
g_\perp^{(v)},g_\perp^{(a)},h_\parallel^{(s)},h_\parallel^{(t)},g_3,h_3\}$
are normalized to satisfy $\int_0^1\!du\, \phi(u) =1$.

The 3-particle distribution amplitudes of the $\rho$ meson are
defined as \cite{rholcda,rhoparameter}
\begin{eqnarray}
\langle 0|\bar u(z) g\widetilde G_{\mu\nu}\gamma_\alpha\gamma_5
  d(-z)|\rho^-(P,\lambda)\rangle &=&
  f_\rho m_\rho p_\alpha[p_\nu e^{(\lambda)}_{\perp\mu}
 -p_\mu e^{(\lambda)}_{\perp\nu}]{\cal A}(v,pz)
\nonumber\\ &&{}
+ f_\rho m_\rho^3\frac{e^{(\lambda)}\cdot z}{pz}
[p_\mu g^\perp_{\alpha\nu}-p_\nu g^\perp_{\alpha\mu}] \widetilde\Phi(v,pz)
\nonumber\\&&{}
+ f_\rho m_\rho^3\frac{e^{(\lambda)}\cdot z}{(pz)^2}
p_\alpha [p_\mu z_\nu - p_\nu z_\mu] \widetilde\Psi(v,pz)\,,\nonumber
\\
\langle 0|\bar u(z) g G_{\mu\nu}i\gamma_\alpha
  d(-z)|\rho^-(P)\rangle &=&
  f_\rho m_\rho p_\alpha[p_\nu e^{(\lambda)}_{\perp\mu}
  - p_\mu e^{(\lambda)}_{\perp\nu}{\cal V}(v,pz)
\nonumber\\&&{}
+ f_\rho m_\rho^3\frac{e^{(\lambda)}\cdot z}{pz}
[p_\mu g^\perp_{\alpha\nu} - p_\nu g^\perp_{\alpha\mu}] \Phi(v,pz)
\nonumber\\&&{}
+ f_\rho m_\rho^3\frac{e^{(\lambda)}\cdot z}{(pz)^2}
p_\alpha [p_\mu z_\nu - p_\nu z_\mu] \Psi(v,pz)\,,\nonumber
\\
\langle 0|\bar u(z) \sigma_{\alpha\beta}
         gG_{\mu\nu}(vz)
         d(-z)|\rho^-(P,\lambda)\rangle
&=& f_{\rho}^T m_{\rho}^3 \frac{e^{(\lambda)}\cdot z }{2 (p \cdot z)}
    [ p_\alpha p_\mu g^\perp_{\beta\nu}
     -p_\beta p_\mu g^\perp_{\alpha\nu}
     -p_\alpha p_\nu g^\perp_{\beta\mu}
     +p_\beta p_\nu g^\perp_{\alpha\mu} ]
      {\cal T}(v,pz)
\nonumber\\
&&+ f_{\rho}^T m_{\rho}^2
    [ p_\alpha e^{(\lambda)}_{\perp\mu}g^\perp_{\beta\nu}
     -p_\beta e^{(\lambda)}_{\perp\mu}g^\perp_{\alpha\nu}
     -p_\alpha e^{(\lambda)}_{\perp\nu}g^\perp_{\beta\mu}
     +p_\beta e^{(\lambda)}_{\perp\nu}g^\perp_{\alpha\mu} ]
      T_1^{(4)}(v,pz)
\nonumber\\
&&+ f_{\rho}^T m_{\rho}^2
    [ p_\mu e^{(\lambda)}_{\perp\alpha}g^\perp_{\beta\nu}
     -p_\mu e^{(\lambda)}_{\perp\beta}g^\perp_{\alpha\nu}
     -p_\nu e^{(\lambda)}_{\perp\alpha}g^\perp_{\beta\mu}
     +p_\nu e^{(\lambda)}_{\perp\beta}g^\perp_{\alpha\mu} ]
      T_2^{(4)}(v,pz)
\nonumber\\
&&+ \frac{f_{\rho}^T m_{\rho}^2}{pz}
    [ p_\alpha p_\mu e^{(\lambda)}_{\perp\beta}z_\nu
     -p_\beta p_\mu e^{(\lambda)}_{\perp\alpha}z_\nu
     -p_\alpha p_\nu e^{(\lambda)}_{\perp\beta}z_\mu
     +p_\beta p_\nu e^{(\lambda)}_{\perp\alpha}z_\mu ]
      T_3^{(4)}(v,pz)
\nonumber\\
&&+ \frac{f_{\rho}^T m_{\rho}^2}{pz}
    [ p_\alpha p_\mu e^{(\lambda)}_{\perp\nu}z_\beta
     -p_\beta p_\mu e^{(\lambda)}_{\perp\nu}z_\alpha
     -p_\alpha p_\nu e^{(\lambda)}_{\perp\mu}z_\beta
     +p_\beta p_\nu e^{(\lambda)}_{\perp\mu}z_\alpha]
      T_4^{(4)}(v,pz)\,,\nonumber
\\
\langle 0|\bar u(z)
         gG_{\mu\nu}(vz)
         d(-z)|\rho^-(P,\lambda)\rangle
&=& i f_{\rho}^T m_{\rho}^2
 [e^{(\lambda)}_{\perp\mu}p_\nu-e^{(\lambda)}_{\perp\nu}p_\mu] S(v,pz)\,,
\nonumber\\
\langle 0|\bar u(z)
         ig\widetilde G_{\mu\nu}(vz)\gamma_5
         d(-z)|\rho^-(P,\lambda)\rangle
&=& i f_{\rho}^T m_{\rho}^2
 [e^{(\lambda)}_{\perp\mu}p_\nu-e^{(\lambda)}_{\perp\nu}p_\mu]
  \widetilde S(v,pz)\,.
\end{eqnarray}
The distribution amplitudes ${\cal A}$, ${\cal V}$ and ${\cal T}$ are of
twist-3 and the other 3-particle distribution amplitudes are of twist-4.


\end{document}